# Observation of topology transition in Floquet non-Hermitian skin effects in silicon photonics

Zhiyuan Lin[1], Wange Song[1, *], Li-Wei Wang[2], Haoran Xin[1], Jiacheng Sun[1], Shengjie Wu[1], Chunyu Huang[1], Shining Zhu[1], Jian-Hua Jiang[2, 3, 4, *], and Tao Li[1, *]

[1]*National Laboratory of Solid State Microstructures, Key Laboratory of Intelligent Optical Sensing and Manipulations, Jiangsu Key Laboratory of Artificial Functional Materials, School of Physics, College of Engineering and Applied Sciences, Nanjing University, Nanjing, 210093, China*

[2]*School of Physical Science and Technology & Collaborative Innovation Center of Suzhou Nano Science and Technology, Soochow University, 1 Shizi Street, Suzhou, 215006, China*

[3]*Suzhou Institute for Advanced Research, University of Science and Technology of China, Suzhou, 215123, China*

[4]*School of Physics, University of Science and Technology of China, Hefei, 230026, China*

[*]*e-mail: songwange@nju.edu.cn, jhjiang3@ustc.edu.cn, taoli@nju.edu.cn*

**Abstract:**

Non-Hermitian physics has greatly enriched our understanding of nonequilibrium phenomena and uncovered novel effects such as the non-Hermitian skin effect (NHSE) that has profoundly revolutionized the field. NHSE is typically predicted in systems with nonreciprocal couplings which, however, are difficult to realize in experiments. Without nonreciprocal couplings, the NHSE can also emerge in systems with coexisting gauge fields and loss or gain (e.g., in Floquet non-Hermitian systems). However, such Floquet NHSE remains largely unexplored in experiments. Here, we realize the Floquet NHSEs in periodically modulated optical waveguides integrated on a silicon photonics platform. By engineering the artificial gauge fields induced by the periodical modulation, we observe various Floquet NHSEs and unveil their rich topological transitions. Remarkably, we discover the transitions between the normal unipolar NHSEs and an unconventional bipolar NHSE which is accompanied by the directional reversal of the NHSEs. The underlying physics is revealed by the band winding in complex quasienergy space which undergoes a topology change from isolated loops with the same winding to linked loops with opposite windings. Our work unfolds a new route toward Floquet NHSEs originating from the interplay between gauge fields and dissipation effects and offers fundamentally new ways for steering light and other waves.

Most systems and materials contain dissipation effects. In situations where the underlying physics can be described by a Hamiltonian, the Hamiltonian will become non-Hermitian with dissipation effects. The study of non-Hermitian Hamiltonians has led to the discovery of many intriguing effects, such as exceptional points [1-5] and nonorthogonal eigenstates [6]. In lattice systems, one particularly interesting phenomenon is the NHSE where an extensive number of eigenstates are localized at the boundary due to non-Hermitian effects. A direct consequence of NHSE is the breakdown of the fundamental notion of the Brillouin zone which must be replaced by the generalized Brillouin zone (GBZ) [7-9] to correctly describe the physical properties of finite systems. Moreover, NHSEs profoundly change the topological bulk-boundary correspondence, leading to rich non-Hermitian topological physics [7,10-25].

Meanwhile, NHSEs can also be regarded as topological effects, of which the topological invariant is the winding number of the complex energy band during its evolution in the Brillouin zone. While a normal unipolar NHSE is manifested in such band winding as a single loop with a finite winding number, a bipolar NHSE exhibits two linked loops with opposite winding numbers, giving rise to distinctive features such as bipolar skin localization and Bloch points [16,26]. In periodically driven systems (i.e., Floquet systems), the artificial gauge fields induced by the periodic driving play pivotal roles in both the NHSEs and the topological properties [27-44], leading to intriguing phenomena at the interface between non-Hermitian and topological physics such as the hybrid skin-topological effects [27,28]. Notably, in Floquet systems, NHSEs can be triggered solely by the on-site loss or gain [27,40-44], whereas the NHSEs in normal systems often rely on nonreciprocal couplings. Furthermore, Floquet non-Hermitian systems also give rise to the discovery of topological lasers [45] and other interesting effects [46-48].

Despite considerable efforts have been made, the experimental realization of Floquet NHSE and its dynamics in integrated photonics remains a big challenge due to the difficulties in simultaneous accurate controls of the gain/loss and Floquet engineering in real optical systems. Here, we propose an efficient approach to realize such Floquet NHSEs through engineering the interplay between the artificial gauge fields and the loss in coupled optical waveguides that are integrated on a silicon photonic chip. Via this approach, we successfully realize and observe various Floquet NHSEs as well as their topological transitions. In particular, we discover a unipolar-bipolar transition in the NHSEs by tuning the modulation periodicity of the width-varying silicon waveguides deposited with loss elements (i.e., chromium) [49]. Remarkably, the bipolar Floquet NHSE state provides a passage for the transition between two unipolar Floquet NHSEs with opposite skin directions. In the complex quasienergy space, the transitions between these Floquet NHSEs are manifested as the change of the topology in the band winding: going from isolated loops with finite winding numbers in the unipolar Floquet NHSEs to linked loops with opposite winding numbers in the bipolar Floquet NHSE.

We start with a one-dimensional lattice of coupled optical waveguides with $N$ unit cells, as shown in Fig. 1(a). Each unit cell (marked by the dot-dashed box) contains three sublattice sites labeled as "A", "B", and "C", respectively. The on-site potential of sublattice site $\xi$ ($\xi \in \{A,B,C\}$) in the $n$th unit cell is $V_{n,\xi}$. The nearest-neighbor couplings $\kappa_{1,2,3}$ are time-dependent and subject to artificial gauge modulations (i.e., $\kappa_m = \kappa\exp[i\theta_m]$ ($m=1,2,3$) and the phases $\theta_m$ are time-dependent). There is an inhomogeneous dissipation distribution in the unit cells, without loss of generality, we set $V_{n,A}=V_{n,B}=0$, while the site C is lossy, $V_{n,C}=i\gamma$. Such a model can be realized in an integrated photonic waveguide system after treating the time dimension as the propagation distance, i.e., $t \to z$ (see **Supplementary Material S1** for more details) [50,51]. In the tight-binding approximation, the $z$-dependent Hamiltonian of the model can be described as

$$H(z) = \sum_{n=1}^{N}\left[\kappa_1(z)|n,A\rangle\langle n,B| + \kappa_2(z)|n,B\rangle\langle n,C| + \text{H.c.}\right] \\ + \sum_{n=1}^{N-1}\left[\kappa_3(z)|n,C\rangle\langle n+1,A| + \text{H.c.}\right] + \sum_{n=1}^{N}\sum_{\xi\in\{A,B,C\}} V_{n,\xi}|n,\xi\rangle\langle n,\xi| \quad (1)$$

where we use the Dirac's notation for quantum mechanics for convenience. For instance, $|n,\xi\rangle$ denotes the state on the sublattice site $\xi$ in the $n$th unit cell. Here, we consider a sinusoidal modulation $\theta_m(z) = \mathcal{V}_m\cos(\Omega z + \varphi_m)$ with the initial phases $\varphi_m = 2m\pi/3$ where $\mathcal{V}_m$ is the amplitude, $\Omega = 2\pi/p$ is the modulation frequency, and $p$ is the modulation period. In our system, the coupling phases $\theta_m$ are realized by engineering the $z$-dependent scalar gauge potentials that control the propagation constants on each sublattice site (i.e., the phases $\Phi_A$, $\Phi_B$, and $\Phi_C$ on each waveguide, see Fig. 1(b)).

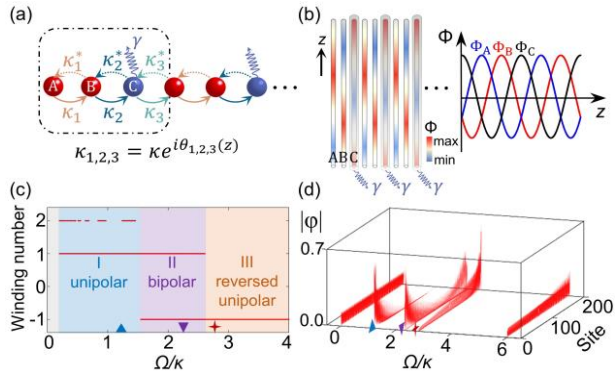

FIG. 1. Floquet NHSE: its realization and topology transition. (a) Schematic of the model. A unit-cell (marked by the dotted box) has three sites (labeled as A, B, and C). Here $\kappa_1$, $\kappa_2$, and $\kappa_3$ are the gauge-modulated coupling coefficients. A and B are lossless and C is lossy. (b) The schematic of the artificial gauge potentials induced by modulating the width of the waveguides (left panel) and the illustration of the gauge potentials on the three waveguides versus the coordinate $z$ (right panel). (c) Winding numbers of band set I as a function of modulation frequency, showing the transition between three Floquet NHSE phases, i.e., the unipolar, bipolar, and reversed unipolar NHSEs

(labeled as blue, purple, and orange, separately). (d) Eigenstate profiles for different modulation frequencies. From left to right, $\Omega/\kappa$=0.0, 1.2, 2.3, 2.8, and 6.0. The upper triangle, down triangle, and the star label the cases of the unipolar, bipolar, and the reversed unipolar NHSEs, respectively.

For our system with periodic modulation, the Floquet Hamiltonian $H_F$ is obtained through the Floquet theory [52,53] (see **Supplementary Material S2**). This Hamiltonian $H_F$ gives the spectrum of quasienergy, $\varepsilon$. The most intriguing property here is the emergence of the Floquet NHSE and its rich topology transition as illustrated in Figs. 1(c)-(d). The Floquet NHSE can be realized without nonreciprocal couplings and its topology can be efficiently tuned by the modulation frequency $\Omega$. In doing so, we can achieve a transition from the unipolar NHSE toward the left boundary, to a bipolar NHSE, and then to a reversed unipolar NHSE toward the right boundary. Remarkably, the transition process is achieved by tuning a single parameter $\Omega$ (or equivalently the modulation period $p$).

The topology of the NHSE can be characterized by the winding number of each band (as derived from $H_F$) in the complex energy space under the periodic boundary condition (PBC) [54,55]

$$w(E) = \frac{1}{2\pi i} \oint_{BZ} \partial_k \ln \det\left[H_F(k) - E\right] dk, \qquad (2)$$

where the integration traverses the first Brillouin zone (BZ). When the spectrum of $H_F$ forms a loop, the above winding number is a topological invariant for any energy within the loop. This winding number characterizes the direction of the exponential decay of the eigenstate wavefunctions associated with the loop and thus determines the direction of the NHSE. In our system, we have three energy bands and the spectrum can be complicated (see Fig. 2(a)). At $\Omega/\kappa$=0, the spectrum does not form any loop and there is no NHSE (Fig. 2(b)(i)). At $\Omega/\kappa$=1.2, the spectrum has three loops which all have the winding number of 1 (Fig. 2(b)(ii)), yielding the NHSE toward the left boundary which is the signature of the unipolar phase I (0.2<$\Omega/\kappa$<1.54). At $\Omega/\kappa$=2.3, two of the three loops become twisted loops, signifying a topology transition, whereas the third loop's topology remains the same (Fig. 2(b)(iii)). Each twisted loop consists of two parts with opposite winding numbers. This corresponds to the region 1.54<$\Omega/\kappa$<2.6, which gives the bipolar phase II. In this phase, the NHSE becomes bipolar, i.e., about half of the eigenstates have wavefunctions localized toward the left boundary, while the other half of the eigenstates are localized toward the right boundary. At $\Omega/\kappa$=2.8, the two twisted loops become simple loops, but the winding number is switched to -1. Thus, this phase (i.e., phase III in the region $\Omega/\kappa$>2.6) corresponds to the unipolar NHSE towards the right boundary (denoted as the reversed unipolar phase). In the above process, the unipolar NHSE switches direction through the bipolar NHSE, yielding rich topology transitions.

It should be noted that the three loops correspond to the three energy bands. As two of these bands may overlap with each other, we denote them as band set I, while the remaining band set is denoted as band set II (see Fig. 2(a)). The topology transition studied here takes place in band set I which is symmetric with respect to Re($\varepsilon$)=0 (The band set II also has topology transitions but not in the same region; see **Supplementary Video 1**). As the two loops in band set I have the same winding number, we shall focus only on the loop which is mainly in the region Re($\varepsilon$)<0. The winding number of such a loop is presented in Fig. 1(c).

Several remarks are in order. First, the strength of the NHSE is determined by the area of the loop. For instance, at very large $\Omega/\kappa$, although the two loops of band set I have winding number $\mathcal{W}$ =-1, the areas of the loops are very small, therefore the NHSE is very weak (see Fig. 1(d) for the eigenstates wavefunctions at $\Omega/\kappa$=6.0). Second, one may notice in Fig. 1c that at some $\Omega/\kappa$, there are two winding numbers: $\mathcal{W}$ =2 and 1. This is because the evolution of the loops is much more complicated than illustrated in Fig. 2 (see such evolution in **Supplementary Video 1**). The winding numbers are calculated for all open-boundary-condition (OBC) eigen-energies enclosed by the loops. Therefore, there can be multiple values of the winding numbers for each set of parameters. Nevertheless, both winding numbers 2 and 1 give rise to the same NHSE towards the left boundary which is consistent with the calculated wavefunctions. Finally, the winding numbers at $\Omega/\kappa$<0.2 are not studied in this work because in this regime the quasienergy bands can be scrambled and the winding numbers can be noninteger. In this region, the calculated eigenstates wavefunctions indicate that the system has no noticeable NHSE (e.g., see Fig. 1(d) and Fig. 2(c)(i)).

To confirm the above phenomena, we demonstrate the dynamics of the light field in the coupled optical waveguide arrays along the *z* direction. The initial excitation is via site B at the center of the waveguide array. Such excitation is chosen because the site B has substantial overlap with the eigenstates of band set I. Consequently, the light dynamics predominantly capture the properties of the eigenmodes of band set I. It should be mentioned that initial excitation at the lossy waveguide C excites mainly the eigenmodes of band set II, which is not considered here. The light propagation dynamics in the optical waveguide system are simulated using coupled-mode theory. As shown in Fig. 2(c)(i), the dynamics at $\Omega/\kappa$=0.0 is consistent with the conventional quantum walk on a discrete lattice, showing no feature of NHSE but instead with a broadened peak at the center of the system. At $\Omega/\kappa$=1.2 (see Fig. 2(c)(ii)), as indicated by the dashed arrow, the NHSE toward the left boundary is notable (Here, the left and right directions are defined in the view of the forward going photons). At $\Omega/\kappa$=2.3, as shown in Fig. 2(c)(iii), there are salient features of the bipolar NHSE. In particular, when compared with the case with $\Omega/\kappa$=0.0, the light wave amplitude at the system center is suppressed, which indicates the bipolar NHSE. At $\Omega/\kappa$=2.8, the light propagation dynamics in Fig. 2(c)(iv) shows clear features of the NHSE toward the right boundary, indicating the reversal of the direction of the NHSE. At significantly large $\Omega/\kappa$ (e.g., $\Omega/\kappa$=6.0), the NHSE is effectively suppressed, and the wave dynamics are similar to that in Fig. 2(c)(v). The wave dynamics for other $\Omega/\kappa$ are also shown in **Supplementary Video 2**. These simulation results are consistent with the prediction from the eigen-spectrum and the winding numbers, indicating that these effects can possibly be observed in optical experiments

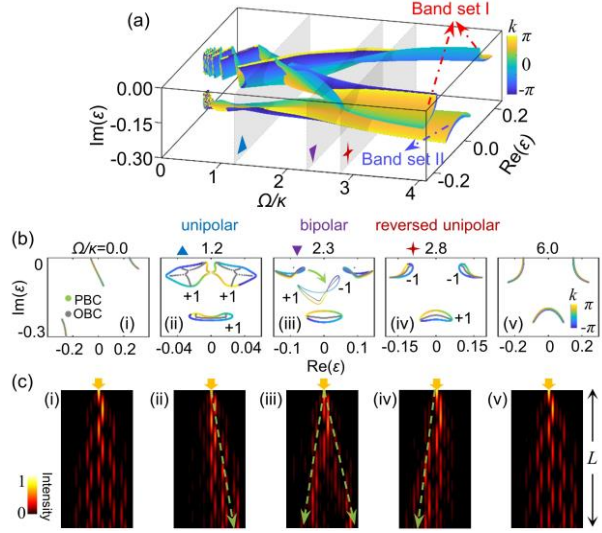

FIG. 2. Quasienergy spectrum, winding numbers, and light propagation. (a) Quasienergy spectrum as a function of the modulation frequency under PBC. (b) Quasienergy spectrum under OBC and PBC for different modulation frequencies $\Omega/\kappa=0.0$ (i), 1.2 (ii), 2.3 (iii), 2.8 (iv), and 6.0 (v). (c) Light propagation dynamics in optical waveguide lattices with 16 sites for $\Omega/\kappa=0.0$ (i), 1.2 (ii), 2.3 (iii), 2.8 (iv), and 6.0 (v). The intensity is normalized to 1 at every $z$ for clearer presentation. Here $\mathcal{V}_m=\nu/\Omega$, $\nu/\kappa=2.0$, $\gamma/\kappa=2.2$, and $L=2L_0$, where $L_0=2\pi/(1.2\kappa)$.

To observe the predicted NHSEs in silicon photonics, we consider an optical lattice comprised of 16 Si-waveguides on the sapphire substrate with air cladding (Fig. 3(a)). The waveguide height is $h=220$ nm and the waveguide spacing is $g=150$ nm. The waveguide width is periodically modulated in a sinusoidal way with averaged width $w=400$ nm and the modulation amplitude $\Delta w=20$ nm. The modulation period $p$ is tuned ($p=\infty$ μm (static), 43 μm, 21 μm, 15 μm, and 8 μm) to achieve different NHSE phases. The loss is introduced by coating a layer of chromium (Cr) with width $w_c=200$ nm and thickness $h_c=8$ nm on top of every three Si waveguides [see **Supplementary Material S3**]. The scanning electron microscope (SEM) pictures of experimentally fabricated samples are shown in Fig. 3(b), where the width modulations and deposited Cr can be clearly observed from the zoom-in picture at the bottom panel.

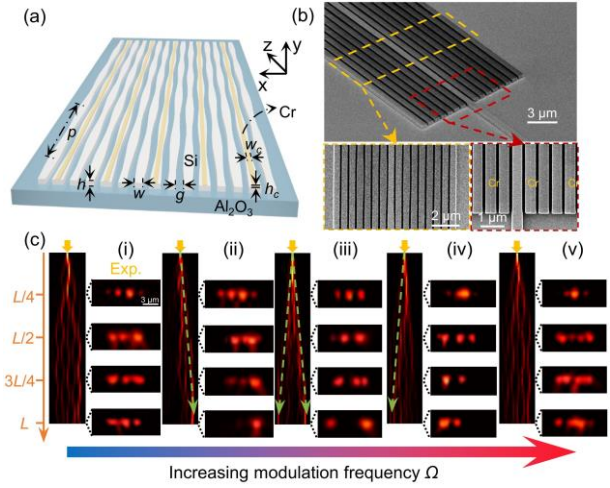

FIG. 3. Realizing Floquet NHSEs in on-chip silicon waveguides. (a) Schematic of the silicon waveguide design. The waveguide width is varied along the propagation direction and Cr is deposited on top of every three silicon waveguides (labeled by yellow). (b) SEM image of one of the experimentally fabricated samples. The bottom panels show the zoom-in views. The waveguide width modulation and Cr strips can be clearly observed (the horizontal-to-vertical scale ratio is 4:1 for clearer presentation). (c) Simulation results for light propagation along the $z$ direction (left) and the experimentally measured light intensities (right) at different propagation lengths ($L/4$, $L/2$, $3L/4$, and $L$, with $L=86$ μm). The modulation periods are (i) $p=\infty$ μm (no modulation), (ii) $p=43$ μm, (iii) $p=21$ μm, (iv) $p=15$ μm, and (v) $p=8$ μm. The intensity is normalized to 1 at every $z$ in the simulation results.

In experiments, we fabricated five types of samples, each with distinct modulation periods, i.e., $p=\infty$ (no modulation), 43 μm, 21 μm, 15 μm, and 8 μm, corresponding to five different cases of non-Hermitian dynamics shown in Fig. 1. To capture various stages of mode evolutions, we fabricated a set of 4 samples with different propagation lengths ($L/4$, $L/2$, $3L/4$, and $L$) for each case. For each sample, a near-infrared laser (1550 nm wavelength) is incident into site B at the center of the waveguide lattices from a grating coupler, and the output intensity distribution was measured by a near-infrared camera through a microscope objective [see **Supplementary Material S3**]. The experimental results are presented in Fig. 3(c), along with simulated light evolutions (using the commercial finite-element software Comsol Multiphysics 5.6) for further verification. When the system is static ($p=\infty$ μm), the light evolution is broadened, as shown in Fig. 3(c)(i). At the appropriate modulation period ($p=43$ μm), light excited from the central waveguide gradually evolves towards one end of the boundary, which is typical of the unipolar skin effect, as shown in Fig. 3(c)(ii). The experimentally captured output results for different propagation lengths demonstrate a gradual unidirectional evolution of the light field towards one end of the boundary with increasing propagation length, aligning closely with the simulation results. By further decreasing the modulation period, e.g., when $p$=21 μm, light excited from the central waveguide eventually evolves towards both ends (Fig. 3(c)(iii)), which is the characteristic of bipolar NHSE. This feature is clearly captured by the experiments, showcasing the light's ultimate tendency to both ends of the boundary, indicating the skin mode transitions. Notably, when $p$=15 μm in Fig. 3(c)(iv), a reversed skin effect becomes evident. The experimental output with different propagation lengths demonstrated consistent results, i.e., the final output spot is at the opposite boundary compared with the NHSE shown in Fig. 3(c)(ii). When the modulation period is much smaller ($p$=8 μm in Fig. 3(c)(v)), the NHSEs become challenging to discern due to the fact that the quasienergy spectra at the PBC and OBC are nearly the same. In this case, the Floquet NHSE is unnoticeable in both experiments and simulations.

The underlying physics of the observed Floquet NHSEs becomes clearer in the Floquet-Hilbert space [52,56,57] [see **Supplementary Material S2**] by mapping the 1D $z$-dependent model into a 2D $z$-independent problem. It is found that non-zero effective magnetic flux will arise in the synthetic static lattice that interacts with loss, and giving rise to nonvanishing currents through the system due to nonzero area of loops in the PBC spectra (Fig. 2), which is the origin of the Floquet NHSEs [16,25]. It is noteworthy that using our design strategy and the controllable Floquet NHSE, an on-chip *Floqurt topological funneling* of light [17] can also be achieved. Such a funnel can steer any light field injected into the structure (irrespective of its shape and input position) toward the system center via the Floquet NHSEs (see more details in **Supplementary Materials S4**).

In summary, we demonstrate the Floquet NHSEs with topology transitions in periodically modulated coupled waveguides on a silicon photonics platform at telecommunication frequencies. By engineering the interplay between the artificial gauge fields induced by the periodic modulation and the optical loss in the waveguides, we observe various Floquet NHSEs and discover their rich transitions. These transitions are manifested as the topology transitions in the band winding in complex quasienergy space. For instance, the transition from a unipolar Floquet NHSE state to a bipolar Floquet NHSE state is accompanied with a transition from a single loop with finite winding number to two linked loops with opposite winding numbers in the complex quasi-energy space. We characterize these Floquet NHSEs via optical measurements and show that they can also be used to realize topological funneling of light via engineering the Floquet NHSE states. These findings unveil a regime where the interplay between gauge fields and dissipation yields fruitful physics and provides new principles for manipulating light and other waves, inspiring future research on exploring various non-Hermitian phenomena such as wave self- healing [12] and self-acceleration [58] in the same silicon photonics platform.

**Acknowledgments**

The authors acknowledge the financial support from The National Key R&D Program of China (No. 2023YFA1407700), National Natural Science Foundation of China (Nos. 12174186, 12204233, 62288101, 92250304, 62325504, and 12125504). Jian-Hua Jiang thanks the support from the CAS Pioneer Hundred Talents Program and the Gusu Leading Innovation Scientists Program of Suzhou City. Tao Li thanks the support from Dengfeng Project B of Nanjing University.